# An Optimized Decision Tree-Based Framework for Explainable IoT Anomaly Detection

**Ashikuzzaman**
Dept. of CSE
University of Barishal
Barishal, Bangladesh
ashikuzzaman.cse7.bu@gmail.com

**Md. Shawkat Hossain**
Dept. of CSE
University of Barishal
Barishal, Bangladesh
shawkat.cse7.bu@gmail.com

**Jubayer Abdullah Joy**
Dept. of CSE
University of Barishal
Barishal, Bangladesh
joy.cse5.bu@gmail.com

**Md Zahid Akon**
Dept. of CSE
University of Global Village
Barishal, Bangladesh
akonzahid0713@gmail.com

**Md Manjur Ahmed***
Dept. of CSE
University of Barishal
Barishal, Bangladesh
mmahmed@bu.ac.bd

**Md. Naimul Islam**
Dept. of CSE
University of Barishal
Barishal, Bangladesh
mnaimul20.cse@bu.ac.bd

*Abstract*—The increase in the number of Internet of Things (IoT) devices has tremendously increased the attack surface of cyber threats thus making a strong intrusion detection system (IDS) with a clear explanation of the process essential towards resource-constrained environments. Nevertheless, current IoT IDS systems are usually traded off with detection quality, model interpretability, and computational effectiveness, thus the deployment on IoT devices. The present paper counteracts these difficulties by suggesting an explainable AI (XAI) framework based on an optimized Decision Tree classifier with both local and global importance methods: SHAP values that estimate feature attribution using local explanations, and Morri's sensitivity analysis identifies the feature importance in a global view. The proposed system attains state of art on the test performance with 99.91% accuracy, F1-score of 99.51% and Cohen Kappa of 0.9960 and high stability is confirmed by a cross validation mean accuracy of 98.93% (($=0.0003$)). Efficiency is also enhanced in terms of computations to provide faster inferences compared to those that are generalized in ensemble models. SrcMac has shown as the most significant predictor in feature analyses according to SHAP and Morris methods. Compared to the previous work, our solution eliminates its major drawback lack because it allows us to apply it to edge devices and, therefore, achieve real-time processing, adhere to the new regulation of transparency in AI, and achieve high detection rates on attacks of dissimilar classes. This combination performance of high accuracy, explainability, and low computation make the framework useful and reliable as a resource-constrained IoT security problem in real environments.

## I. INTRODUCTION

The Internet of Things (IoT) already transforms more industries by connecting billions of dissimilar devices. Statista says that the amount of connected IoT devices worldwide will reach over 30 billion by 2030 [1]. As this development leads to innovative solutions in healthcare, transport, and smart cities, it also comes with a dramatically increased attack surface. This predisposes IoT scenario to various attacks, including lack of sufficient computing resources, secured communication protocols, and strict security measures [2] [3].

Intrusion Detection Systems (IDS) have emerged to be unavoidable in IoT networks protection [4]. Nevertheless, current IDS strategies can be marred with various drawbacks, such as the low precision to handle high-dimensionality and imbalanced data, high-computational costs that cannot be afforded by a resource-limited IoT device, and their limited interpretability, which makes their predictive study least trusted or verified. Such shortfalls show an urgent need to develop an IDS that is lightweight and correct as well as transparent and explainable.

To address these issues, we present one of the first and interpretable IoT intrusion detection systems that incorporates optimized machine learning and explainable AI (XAI). To start their proposed methodology, a complete pre-processing pipeline will be used to normalize data, deal with missing values, and reduce the imbalance of classes. We next use Recursive Feature Elimination (RFE) and cross-validation (CV) in order to find the set of features that explain our data the most, which is correlated with improvement in dimensionality and generalization of a model. Decision Tree (DT) classifier is utilized due to a balance between efficiency and explainability of this model. To further improve interpretability and instill confidence into the decisions of the model, we use XAI such as SHAP (SHapley Additive exPlanations) and the Morris sensitivity method.

The major contributions of this research are summarized as follows:

- A preprocessing pipeline is developed for IoT traffic, including normalization, noise reduction, and class imbalance handling.
- Recursive Feature Elimination with cross-validation is applied to select key features, enhancing accuracy and efficiency.

- A Decision Tree classifier is optimized for high accuracy and low computational cost, making it suitable for real-time IoT deployment.
- Explainable AI techniques (SHAP and Morris sensitivity analysis) are integrated to improve transparency and trust in model decisions.
- The proposed framework is validated on benchmark IoT intrusion datasets, demonstrating improved accuracy, interpretability, and reduced resource consumption.

The rest of the paper is organized as follows: Section II reviews related work. Section III describes the proposed approach. Section IV presents experiments and results. Section V concludes the paper.

## II. Related Work

The recent developments of anomaly-based intrusion detection system (A-IDS) to IoMT and IoT integrate deep learning, explainability, and ensemble techniques to overcome the ordeal issues. The RCLNet proposed by Shaikh [5] has the possibility of maximizing the utilization of features selected by Random Forest and including them as part of a CNN-LSTM and Self- Adaptive Attention Layer and reaches a 99.78% of accuracy on the WUSTL-EHMS-2020 dataset but their scalability and implementation in real-time applications has not been tested. Hosain et al. [6] created XAI-XGBoost with recursive feature elimination and feature importance methods SHAP and LIME, achieving 99.22 per cent accuracy, yet the computational costs and generalization are still unknown. Alnasrallah et al. [7] integrated Information Gain and Recursive Feature Elimination into a deep autoencoder and DNN classifier and achieved 99.93% accuracy and less training time, though its applicability to resource-constrained IoMT devices has to be determined. An active learning framework was recently published by Zakariah et al. [8] that minimizes labeling work whilst retaining 99.7% accuracy on UNSW-NB15 but the limitation of its application scale and in streaming settings is not discussed. In the case of unsupervised detection, the accuracy obtained in BoT-IoT was of 99.99 by Ayad et al. on a one-class autoencoder based DNN and a one-class DNN, but the data drift robustness is not clear. Bayesian hyperparameter optimization by Lai et al. [9] can improve the performance of ensemble models by up to 30 percent in F1- score, although the computational cost and even the possibility of deployment at edges are problematic. Lastly, Sathyabama and colleagues [10] connected deep neural networks with blockchain to detect anomalies and inopportune logging to achieve 99.18% accuracy; however, blockchain overheads and employability are still problematic. Alsalman study involves a robust collection of data that is used to study the behaviors of the networks in the Internet of Medical Things (IoMT) [11]. Authors used ensemble models consisting of Multi- Layer Perceptrons (MLP) based models to detect anomaly, and through FusionNet, they got astonishing results with Fusion- Net reaching 99.5% accuracy, 99% precision, 99.8% recall, and 99.7% F1 score. Nevertheless, the issue with overfitting is possible, as well as the thorough cross-validation is required.

In addition, the application of Explainable AI (XAI) solutions may increase transparency and inspire trust in the mode of operation of the model, especially in the case of sensitive applications in the healthcare field. Haque et al. [12] address the challenge of increasing cyber threat of IoT networks due to excessive imbalanced data. They use hybrid sampling methods in order to improve the accuracy of the detection evaluation and test various methods of machine learning such as Random Forest and Soft Voting. Random Forest model had the high value of Kappa (0.9903), 99.61 percent test accuracy, and AUC (0.9994), and Soft Voting had the value of 99.52 percent accuracy and 0.9997 AUC. The research is however not explainable (XAI) and there is no mention of overfitting that can interfere with the applicability of the model to real- life situations.

**Research Gaps:** Even though they have a high level of accuracy, IoT anomaly techniques have explainability challenges brought about by the computational overhead and limited real-time possibility. Its ability to generalize on a wide and changing set of data is unexplored, and it is mostly validated on tiny benchmarks. There are limitations on scalability, resources constraints, and edges devices to implement on resource-constrained and edge devices especially deep learnng ensemble and blockchain-based methods. Active learning would lessen the amount of labeling work but is not tested in large-scale streaming contexts. One-class models that do not require a label are vulnerable to the performance decreases induced by the concept of drift, and therefore it emphasizes the necessity of adaptive, resourceful, and scalable IDS systems.

## III. Methodology

This section outlines the step-by-step approach adopted for anomaly detection in IoT networks. The methodology encompasses data collection, preprocessing, model development, training, and evaluation. Figure 1 shows the overall workflow diagram

### A. Data Collection

The dataset available in the research was retrieved in Washington University in St. Louis (WUSTL), which contains three-class networks: Normal, Spoofing, and Data Alteration [13] Answers The number of the samples are 16,318, 14,272 as Normal, 1,124 as Spoofing, and 922 as Data Alteration. The samples are found to have 41 variables, four of them being categorical and the rest (thirty-seven) numerical. These attributes intercept different traffic attributes of a network that are needed in the process of intrusion detection.

### B. Preprocessing

Duplicate records were rid off the data to guarantee data integrity. Label encoding was used to convert categorical variables into numerical form that can be used in machine learning algorithms. Continuous features that had outliers were identified using the interquartile range (IQR) technique and their outlier filled with features means to reduce their impact on the modeling outcome. Last but not least, to normalize the scale of the numerical features, z-score normalization was applied to them to ensure better training stability and faster convergence.

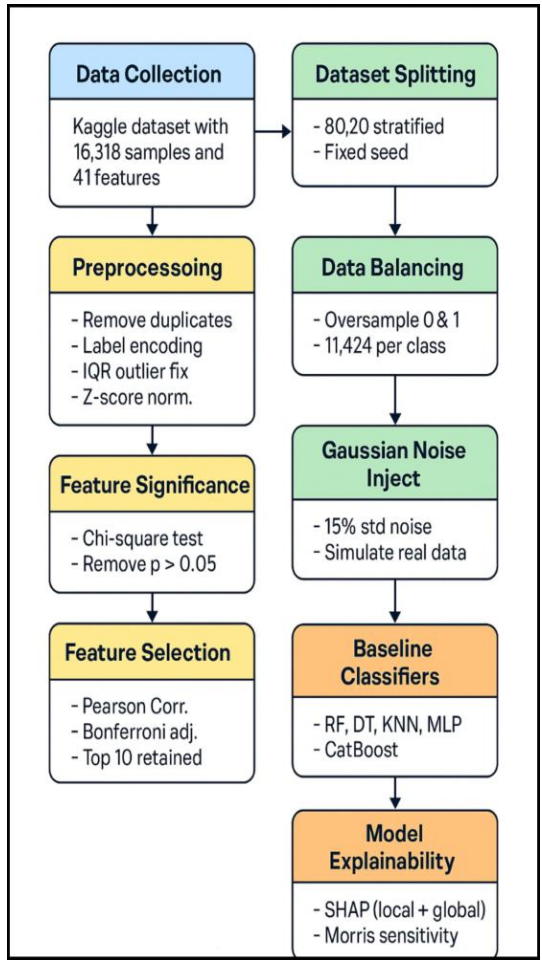

Fig. 1: Overall workflow diagram for anomaly detection in IoT networks.

### C. Feature Significance

The chi-square test of independence, ($\chi^2$), was used to perform feature evaluation to quantify the relationship between a given feature and the target variable, *Attack Category*. The chi-square statistics and associated p-values were computed and the feature whose p-value was greater than 0.05 were dropped to make the dataset less complex to improve the effectiveness of the model. This methodology was in order to avoid retaining those features that were not very relevant to the target variable.

### D. Feature Selection

Each feature was Pearson correlation correlated with the target variable and missing values were removed. Multiple comparisons (Bonferroni correction) were controlled by keeping the features with adjusted p-value less than 0.05. Table I shows the ten most important features sorted by their absolute correlation.

TABLE I: Top 10 Features by Pearson Correlation

| Feature | Correlation | Adjusted p-value |
| --- | --- | --- |
| SrcAddr | 0.85 | <0.0001 |
| DstAddr | -0.78 | <0.0001 |
| Sport | 0.75 | <0.0001 |
| DstMac | -0.70 | <0.0001 |
| SrcMac | 0.68 | <0.0001 |
| PacketLength | -0.65 | <0.0001 |
| FlowDuration | 0.60 | <0.0001 |
| AvgPacketSize | -0.58 | <0.0001 |
| Protocol | 0.55 | <0.0001 |
| Timestamp | -0.52 | <0.0001 |

### E. Dataset Splitting

The preprocessed data was divided into the training and testing set with 80:20 and a fixed seed random to make it reproducible. They had 12,054 training samples, which were divided into three categories of attacks, which include, 11,424 normal, 895 spoofing, and 735 data alteration. The distribution of samples used in the test set was 3,264 samples, of which are 2,848 normal, 229 spoofing, and 187 data alteration. Such stratified division maintained the distribution of the classes, balanced the data to be used in model training and evaluation (see Table II).

TABLE II: Class Distribution in Training and Test Sets

| Attack Category | Training Set | Test Set | Total |
| --- | --- | --- | --- |
| Normal (2) | 11,424 | 2,848 | 14,272 |
| Spoofing (1) | 895 | 229 | 1,124 |
| Data Alteration (0) | 735 | 187 | 922 |
| **Total** | 12,054 | 3,264 | 15,318 |

*1) Data Balancing:* To solve the issue of class imbalance in training set, minority classes related to attack class 0 and 1 were oversampled by random oversampling to the same number of classes with the majority class category 2. In particular, categories 0 and 1 sample were resampled randomly with replacement up to the number of 11,424 instances (used by the majority category). The resulting balanced training-data consisted of 34 272 samples and each category was distributed in portions of 12023. The data were shuffled to enhance randomness in them. The distribution of classes after balancing has been summarized in Table III.

TABLE III: Class Distribution After Training Set Balancing

| Attack Category | Number of Samples |
| --- | --- |
| Normal (2) | 11,424 |
| Spoofing (1) | 11,424 |
| Data Alteration (0) | 11,424 |
| **Total** | 34,272 |

## F. Baseline Classifier

To increase model robustness, a Gaussian noise was applied on the balanced feature set. Particularly, one injected 0-mean noise, whose standard deviation was taken to be equal to 15 percent of the original standard deviation of each feature to simulate the real-world variability in data. Several basics classifiers (Random Forest, Decision Tree, K-Nearest Neighbors, Multi-Layer Perceptron, CatBoost) were trained on this data and tested. Reproducibility of the experiments was guaranteed because of a fixed random seed. Their primary hyperparameters during training are described in Table IV.

TABLE IV: Baseline Classifiers and Key Hyperparameters

| Classifier | Key Parameters |
| --- | --- |
| Random Forest | 100 trees, max depth 10, min samples split 4 |
| Decision Tree | max depth 10, min samples split 4 |
| K-Nearest Neighbors | 5 neighbors, uniform weights |
| Multi-Layer Perceptron | 1 hidden layer (128 units), 1000 max iterations, $\alpha = 0.001$ |
| CatBoost | 500 iterations, learning rate 0.1, depth 6, $L_2$ Leaf regularization 10 |

## G. Model Explainability

To make it easier to interpret, explainable Artificial Intelligence (XAI) methods were applied into the framework. The SHAP was used to measure the contribution of each feature to the individual predictions by having both global and local interpretability. Also, Morri's sensitivity analysis was used to determine the overall effect of the input features on model output, including nonlinear effect and interaction. This two-fold strategy provides thorough transparency, which would be useful in terms of trust and accountability to the intrusion detection system.

## IV. RESULTS AND DISCUSSION

Table V presents the comparative study of the suggested optimized Decision Tree, Random Forest, K-Nearest Neighbors (KNN), Multi-layer Perceptron (MLP), and CatBoost machine learning classifiers when used with the noisy IoT anomaly detection dataset. Accuracy, precision, recall, F1-score, and Cohen kappa were used to judge performance on both the training as well as testing sets. The suggested idealized Decision Tree classifier proved to be the best with the best testing precision (99.91%) and F1-score (99.51%), the best training measures were also stable. This shows that this has powerful generalization abilities in the event of label noise, which is the typical feature of irreal world IoT sensor data. Random Forest and KNN demonstrated quite a robust effect with their testing F1-scores of more than 99.0%, not to mention the fact that MLP and CatBoost were also competitive in all the metrics examined. These findings show the robustness of the classical and ensemble learning algorithms even in noisy domains. Noteworthy, selecting the model in the IoT systems should regard not only predictive accuracy but also questioning computational performance. Ensemble models, e.g., Random Forest and CatBoost, are more precise but require increased computational cost, which can be a problem in the deployment to the resource-limited edge devices. Comparatively, the advanced Decision Tree model exhibits high performance in a simple appearance, which makes them an acceptable model for anomaly detection in a distributed Internet of Things. In addition, 5-fold cross-validation was performed to calculate the stability of the models, which again confirmed that the proposed optimized Decision Tree obtained the most constant results when using various data splits.

Another measure of model reliability is given in the table above, Table VI, which gives the mean value of cross-validation accuracy and standard deviation over five folds. The optimization result of the proposed Decision Tree had the best consistency, lowest standard deviation of 0.0003 and high mean accuracy of 98.93 percent. Such consistency between data partitions is what is required in noisy and dynamic IoT settings. As much as the mean accuracies skewed towards each other, the standard deviation values were higher in Random Forest and MLP, which implies inconsistent performance. CatBoost despite its competitive accuracy, presented the most significant variability, which could limit its real world deployability. Altogether, these results highlight the suggested optimized Decision Tree as a good tradeoff between precision and stability, and, therefore, the reason why the proposed approach could be well adopted to finding anomalies in IoT efficiently.

To further complete the accuracy and stability analysis, Figure 2 shows the normalized training and inference times of the five classifiers. CatBoost takes the most time to train and the least time to get inferences and KNN trains the fastest but takes the longest to get inferences. Random Forest has a comparatively slow training time and high prediction speed. The suggested optimized Decision Tree is distinguished because it integrates minimal training and inference times, which is why it is one of the most effective in terms of RT anomaly detection in IoT. MLP illustrates a moderate amount of training time and fast inference. This radar chart can briefly illustrate the trade-offs of computational costs of such models, adding clarity to the relative heat of deploying each of them in IoT networks with limited resources.

## A. Model Performance Visualization

In order to evaluate the power and stability of the suggested optimized Decision Tree classifier we examined its confusion matrix, Precision-Recall curves, as well as ROC curves over noisy IoT anomaly data. The confusion matrix (top, Figure 3) indicates that the classification almost does not have errors, except those in Class 1. There is a good balance between the precision and recall values as depicted by the Precision-Recall curves (middle) whereby all classes yielded AUC values that are higher than 0.99. There are a very good sensitivity and specificity reported by the ROC curves (bottom) with AUC scores near to 1.0. The summation of these outcomes ensures the strength and validity of the model in real-time anomaly detection couple with noisy surroundings.

TABLE V: Performance Comparison of Models on Training and Testing Sets

| Model | Training Metrics | | | | Testing Metrics | | | | |
|---|---|---|---|---|---|---|---|---|---|
| | Accuracy | Precision | Recall | F1-score | Accuracy | Precision | Recall | F1-score | Kappa |
| **Proposed Optimized Decision Tree** | 0.9985 | 0.9951 | 0.9940 | 0.9945 | 0.9991 | 0.9947 | 0.9956 | 0.9951 | 0.9960 |
| Random Forest | 0.9946 | 0.9911 | 0.9905 | 0.9908 | 0.9940 | 0.9903 | 0.9908 | 0.9905 | 0.9954 |
| KNN | 0.9942 | 0.9904 | 0.9900 | 0.9902 | 0.9938 | 0.9901 | 0.9906 | 0.9902 | 0.9950 |
| MLP | 0.9939 | 0.9899 | 0.9895 | 0.9897 | 0.9935 | 0.9898 | 0.9903 | 0.9900 | 0.9943 |
| CatBoost | 0.9936 | 0.9892 | 0.9887 | 0.9889 | 0.9931 | 0.9895 | 0.9900 | 0.9897 | 0.9940 |

TABLE VI: Mean Cross-Validation Accuracy and Standard Deviation of Models

| Classifier | Mean CV | Standard Deviation |
|---|---|---|
| Proposed Optimized Decision Tree | **0.9893** | **0.0003** |
| Random Forest | 0.9892 | 0.0014 |
| KNN | 0.9887 | 0.0011 |
| MLP | 0.9889 | 0.0013 |
| CatBoost | 0.9885 | 0.0015 |

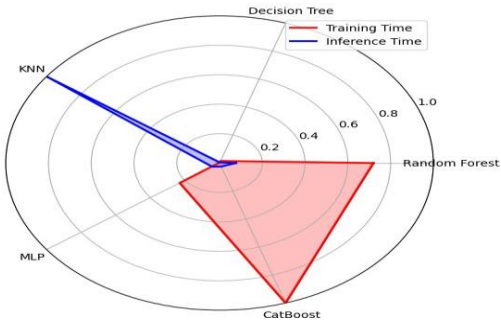

Fig. 2: Radar chart comparing normalized training and inference times of five classifiers on noisy IoT anomaly detection data.

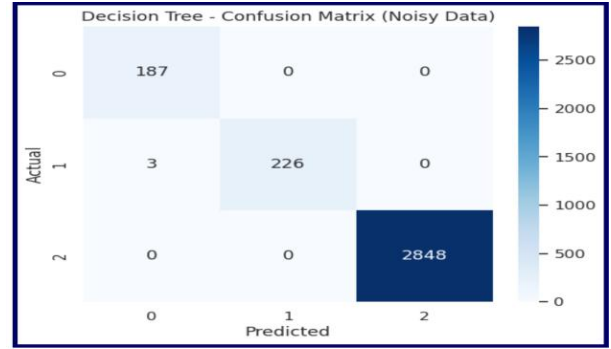
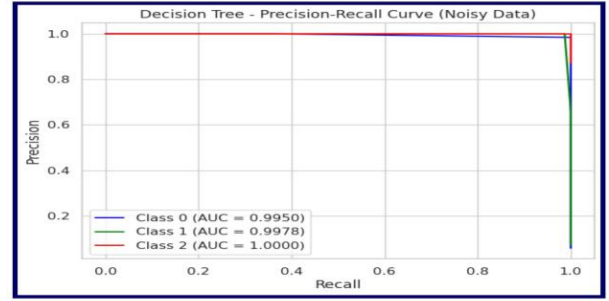
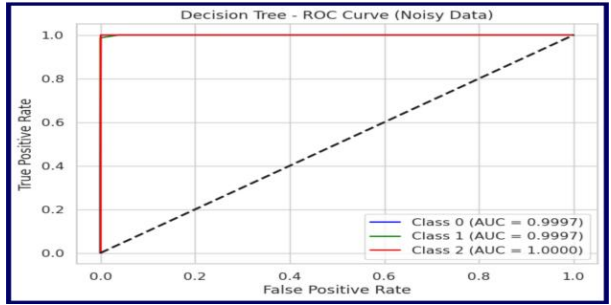

Fig. 3: Confusion matrix (top), Precision-Recall curves (middle), and ROC curves (bottom) for the proposed optimized Decision Tree classifier on noisy IoT anomaly data.

Also, a visualization of the first four depths of the optimized Decision Tree is shown in Figure 4 to reveal how the model makes the decisions. Collectively, these findings demonstrate that the model was highly discriminative, resistant to noise and reliable, and therefore, very appropriate in real time in anomaly detection in the IoT systems.

### B. Explainable AI: Insights and Interpretations

Complementary interpretability of the proposed model is shown in figure 5. The SHAP summary plot (top), validates that SrcMac was the main predictor amongst all the classes, especially Class 2. Also, SIntPkt and DIntPkt play a big role, in particular, when it comes to Classes 0 and 1. Those features with the least SHAP values correspond with those that the Morris technique identifies (bottom) and have little action accordingly. Morris, a global sensitivity analysis method that measures the overall effect of input features by perturbing them and measuring a change in the output identifies SrcMac, SIntPkt and DIntPkt as the most influential features. This can be confirmed by their large * values with great confidence intervals. The metrics, including DstLoad, Rate, Load, and jitter, are not very crucial in the two analyses and hence do not have much weight in the predictions that the models make.

### C. Comparative Analysis with previous studies

Table VII proves that our optimized/enhanced Decision Tree is better than most of the literature reviewed, in that it reports the high accuracy score of 99.91%, standing almost next to the ones reported by the best-performing models. Notably, in contrast to other most-accurate methods, which cannot be explained, our model possesses a decision-making process transparency, making it accessible and explainable in the domain of sensitive IoT applications, complying with the requirements of Explainable AI (XAI). Also, as opposed to the deep learning or ensemble techniques, optimization of Decision Tree consumes much less computational resources, thus it is suitable to be executed on resource-limited IoT devices. Such high accuracy, explainability, and low computational overhead make our solution a viable and stable tool that can be used in real-time anomaly detection on the IoT network.

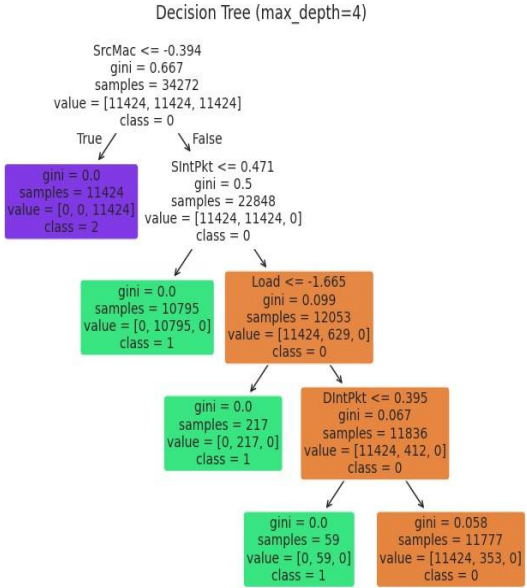

Fig. 4: Visualization of the first four depths of the proposed optimized Decision Tree classifier.

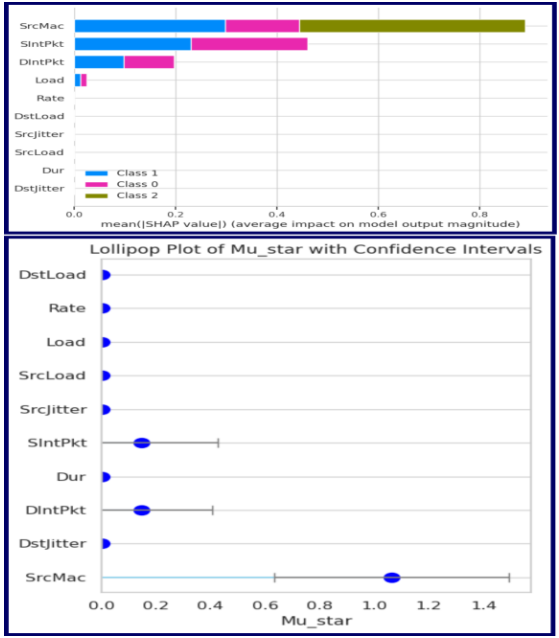

Fig. 5: Top: SHAP summary plot showing feature importance across classes. Bottom: Morri's method lollipop plot illustrating global sensitivity indices (*) with confidence intervals for each feature.

TABLE VII: Comparative Analysis of IoT Anomaly Detection Studies

| Study | Best Model | Accuracy (%) | XAI |
|---|---|---|---|
| [5] | CNN-LSTM | 99.78 | No |
| [6] | XGBoost | 99.22 | Yes |
| [7] | DNN | 99.93 | No |
| [14] | One-Class DNN | 99.99 | No |
| Our Study | Optimized Decision Tree | 99.91 | Yes |

## V. CONCLUSION

This paper suggests an interpretable and computationally lightweight framework of intrusion detection tailored to the IoT setting, which requires efficient, transparent models. The system uses a fine-tuned Decision Tree classifier and SHAP and Morris sensitivity analysis to deliver good performance as well as feature-level explainability. It runs with a test accuracy of 99.91 percent, F1-score of 99.51 percent and Cohen Kappa of 0.9960 which are strong and substantiated by a 98.93 percent cross-validation mean and a very low standard deviation of 0.0003. Some of the main characteristics like *SrcMac*, *SIntPkt* and *DIntPkt* were always found to be extremely influential. The framework also substantially decreases the inference time compared with ensemble baselines, and this allows us to use the framework in real time on the edge. Future efforts will take the form of the integration of adaptive learning processes and interpretable deep models to allow a greater sense of response to adaptive IoT threats. Larger and broader datasets will assist in better generalization, and implementation of the framework in real-world settings of IoT will also be able to test its stability and capacity to be applied in practice.